\input phyzzx
\hsize=417pt 
\sequentialequations
\Pubnum={ EDO-EP-8}
\date={ \hfill January 1997}
\titlepage
\vskip 32pt
\title{ Mass Inflation in Quantum Gravity }
\author{Ichiro Oda \footnote\ddag {E-mail address: 
sjk13904@mgw.shijokyo.or.jp}}
\vskip 12pt
\address{ Edogawa University,                                
          474 Komaki, Nagareyama City,                        
          Chiba 270-01, JAPAN     }                          
%
%
%
%
%
\abstract{ Using the canonical formalism for spherically symmetric black 
hole inside the apparent horizon we investigate the mass inflation in the 
Reissner-Nordstr$\ddot o$m black hole in the framework of quantum 
gravity. It is shown that like in classical gravity the combination of 
the effects of the influx coming from the past null infinity and the 
outflux backscattered by the black hole's curvature causes the mass 
inflation even in quantum gravity.
The results indicate that the effects of quantum 
gravity neither alter the classical picture of the mass inflation nor  
prevent the formation of the mass inflation singularity. } 
\endpage
%
%
%

\def\sp(#1){\noalign{\vskip #1pt}}

%
%
%
%
%
\topskip 30pt
\par
\leftline{\bf 1. Introduction}	
\par
It is well known that a spherically symmetric solution of the Einstein 
equation with an electric charge is Reissner and Nordstr$\ddot o$m 
geometry [1]. Since we expect that any macroscopic body in astronomical 
universe does not possess a net electric charge by the neutralization 
process with the surrounding, a consideration of 
the Reissner-Nordstr$\ddot o$m black hole may seem to be outside the 
realm of reality. 
Nevertheless, a study of the Reissner-Nordstr$\ddot o$m black 
hole leads us to a useful understanding of the structure of the spacetime. This 
is because in spite of its simple geometrical form 
the Reissner-Nordstr$\ddot o$m black hole has two horizons, 
an external event horizon and an internal Cauchy horizon, whose natures 
are also shared by a more realistic Kerr black hole. 

The existence of a region beyond the Cauchy horizon is rather disturbing 
since one cannot predict a future evolution of events in this region 
since the informations which cannot be determined by the initial data 
come from the singularity owing to the timelike character in an 
uncontrollable way. Thus the 
Cauchy horizon would give us a possibility of the violation of the strong 
cosmic censorship hypothesis of Penrose [2].

The Cauchy horizon in both the Reissner-Nordstr$\ddot o$m black 
hole and the Kerr one is unstable against small perturbations in the 
initial data [3]. Namely, the energy-momentum tensor associated with the 
matter fields diverges at the Cauchy horizon since this null surface is 
that of infinite blueshift. It is thus widely believed that the Cauchy 
horizon would become a curvature singularity if the back reaction of the 
matter fields on the geometry is taken account of in a self-consistent 
manner although we do not have a satisfying proof of this conjecture yet 
[3].

A few years ago, Poisson and Israel have proposed an interesting model 
which examines the physical behavior of the Cauchy horizon of the 
charged black hole under the back reaction of the 
matter fields [4]. They have shown that the curvature singularity is 
indeed formed along the Cauchy horizon but this singularity is rather
mild in that the metric tensor is regular while the derivatives of it 
diverge on the Cauchy horizon. Moreover it was shown that this 
singularity is characterized by the exponential divergence of the local 
mass 
function with an advanced time, what is called, mass inflation. This 
interesting phenomenon have been subsequently investigated in more detail 
[5, 6, 7, 8].  

Recently we have constructed 
the canonical formalism of a system with a spherically symmetric black 
hole inside the apparent horizon [9]. Since in this region the Killing 
vector field $\partial \over {\partial t}$ becomes spacelike 
while it does timelike in the exterior region, 
one must foliate the interior region of a black hole by a 
family of spacelike hypersurfaces, for example, $r = const$. As one of 
applications of this canonical formalism, we have investigated the black 
hole radiation and shown that the mass-loss rate by the black hole 
radiation is exactly equal to that evaluated by Hawking in 
the semiclassical approximation [10].

In this article, based on the canonical formalism constructed in 
the previous work [9] we would like to study how quantum gravity would modify 
the classical picture of the mass inflation. Since singular behavior of 
the Cauchy horizon suggests that the classical theory of 
general relativity would be broken down and quantum gravity would play a 
dominant role there, it is very natural to study the mass inflation in the 
framework of quantum gravity.

The article is organized as follows. In section 2, we review a 
construction of the canonical formalism inside the apparent horizon.
In section 3, by using the 
canonical formalism reviewed in section 2 we will make a formalism of 
quantum gravity holding in the vicinity of the apparent horizons. 
In section 4, we apply this formalism for the mass inflation. 
The last section is devoted to conclusion.

\vskip 1cm
\leftline{\bf 2. Review of canonical formalism}	
\par
We begin by a review of our previous work [9] of constructing a canonical
formalism of a spherically symmetric system with a black hole inside the 
horizon (See the references [11, 12] for the canonical formalism outside 
the horizon). The important point is that the radial 
coordinate plays a role of time inside the horizon in the spherically 
symmetric coordinate system. Thus we must take a choice of 
$x^1 = const$ hypersurfaces to slice the spacetime. Later we will take 
the simplest choice $x^1 = r$.  

The four dimensional action which we consider in this section is of the 
form $\footnote\dag {We replace the coefficient of the second term 
${1 \over 4 \pi}$ with ${1 \over 8 \pi}$ in case of the neutral scalar 
field as will be treated in later sections.}$
$$ \eqalign{ \sp(2.0)
S = \int \ d^4 x \sqrt{-^{(4)}g} \bigl[ {1 \over 16 \pi} {}^{(4)}R - {1 \over 
4 \pi} {}^{(4)}g^{\mu\nu} (D_{\mu} \Phi)^{\dag} D_{\nu} \Phi -  {1 \over 
16 \pi e^2} F_{\mu\nu}F^{\mu\nu} \bigr],
\cr
\sp(3.0)} \eqno(1)$$
where $\Phi$ is a complex scalar field, $A_{\mu}$ the electromagnetic field, 
$F_{\mu\nu}$ the corresponding field strength,
and $e$ is the electric charge of $\Phi$. To clarify the four dimensional 
meaning we put the suffix $(4)$ in front of the metric tensor and the 
curvature scalar. We follow the conventions adopted in the MTW textbook 
[13] and use the natural units $G = \hbar = c = 1$. The Greek indices 
$\mu, \nu, ...$ take the values 0, 1, 2, and 3, on the other hand, 
the Latin indices $a, b, ...$ take the values 0 and 1. 

The most general spherically symmetric assumption for the metric is
$$ \eqalign{ \sp(2.0)
ds^2 &= {}^{(4)}g_{\mu\nu} dx^{\mu} dx^{\nu},
\cr
     &= g_{ab} dx^a dx^b + \phi^2 ( d\theta^2 + \sin^2\theta d\varphi^2 ), 
\cr
\sp(3.0)} \eqno(2)$$
where the two dimensional metric $g_{ab}$ and the radial function $\phi$ 
are the functions of only the two dimensional coordinates $x^a$. The 
substitution of (2) into (1) and then integration over the angular 
coordinates $(\theta, \varphi)$ leads to the following two dimensional 
effective action
$$ \eqalign { \sp(2.0)
S &= {1 \over 2} \int \ d^2 x \sqrt{-g} \bigl[ 1 + g^{ab} \partial_a \phi 
\partial_b \phi +  {1 \over 2} R \phi^2 \bigr] 
\cr
&\qquad- \int \ d^2 x \sqrt{-g} \phi^2 g^{ab} (D_a 
\Phi)^{\dag} D_b \Phi - {1 \over 4} \int \ d^2 x \sqrt{-g} \phi^2 F_{ab} 
F^{ab},
\cr
\sp(3.0)} \eqno(3)$$
where 
$$ \eqalign{ \sp(2.0)
D_a \Phi = \partial_a \Phi + i e A_a \Phi.
\cr
\sp(3.0)} \eqno(4)$$

Next let us rewrite the action (3) into the first-order ADM form. As 
mentioned before, we shall take the $x^1$ coordinate as time 
to cover the internal region of a black hole by spacelike hypersurfaces. 
Then the appropriate ADM splitting of (1+1)-dimensional spacetime is 
given by 
$$ \eqalign{ \sp(2.0)
g_{ab} = \left(\matrix{ \gamma  & \alpha \cr
              \alpha & {\alpha^2 \over \gamma} - \beta^2 \cr} \right),
\cr
\sp(3.0)} \eqno(5)$$
and the normal unit vector $n^a$ which is orthogonal to the hypersurface 
$x^1 = const$ reads 
$$ \eqalign{ \sp(2.0)
n^a = ({\alpha \over {\beta \gamma}}, - {1 \over \beta}).
\cr
\sp(3.0)} \eqno(6)$$

After using the various formulae derived in the Ref.[9], the action (3) 
can be written as
$$ \eqalign{ \sp(2.0)
S &= \int d^2x L  =\int d^2x \bigl[ {1 \over 2} \beta \sqrt{\gamma} 
\bigl\{ 1 - (n^a \partial_a 
 \phi)^2 + {1 \over \gamma} \dot \phi^2 - K n^a \partial_a (\phi^2) 
\cr
&\qquad+ {\dot \beta \over {\beta\gamma}} \partial_0 (\phi^2) \bigr\} 
+ \beta \sqrt{\gamma} \phi^2 \bigl\{{ | n^a D_a \Phi |^2 - {1 \over \gamma}
|D_0 \Phi |^2 }\bigr\} + {1 \over 2} \beta \sqrt{\gamma} \phi^2 E^2 \bigr] 
\cr
&\qquad+ \int d^2x \bigl[ {1 \over 2} \partial_a 
 (\beta \sqrt{\gamma} K n^a \phi^2 ) 
- {1 \over 2} \partial_0 ({\dot \beta 
 \over {\sqrt{\gamma}} \phi^2}) \bigr]\hfill,
\cr
\sp(3.0)} \eqno(7)$$
where
$$ \eqalign{ \sp(2.0)
E&= {1 \over \sqrt{-g}} F_{01} = {1 \over \beta \sqrt{\gamma}} 
(\dot A_1 - A_0^{\prime}),
\cr
\sp(3.0)} \eqno(8)$$
the trace of the extrinsic curvature $K = g^{ab} K_{ab}$ is expressed by
$$ \eqalign{ \sp(2.0)
K = - {\gamma^\prime \over {2\beta\gamma}} + {\dot \alpha \over 
{\beta\gamma}} - {\alpha \over {2\beta\gamma^2}} \dot \gamma,
\cr
\sp(3.0)} \eqno(9)$$
and ${\partial \over {\partial x^0}} = \partial_0$ and ${\partial \over 
{\partial x^1}} = \partial_1$ are also denoted by an overdot and a prime, 
respectively.

By taking the variation of the action (7) with respect to the $x^1$ 
derivative of the canonical 
variables $\Phi (\Phi^{\dag}), \phi, \gamma$ and $A_0$ we have the 
corresponding conjugate momenta $p_{\Phi} ({p_{\Phi^{\dag}}}), p_{\phi}, 
p_{\gamma}$ and $p_A$ 
$$ \eqalign{ \sp(2.0)
p_{\Phi} = - \sqrt{\gamma} \phi^2 (n^a D_a\Phi)^{\dag},
\cr
\sp(3.0)} \eqno(10)$$
$$ \eqalign{ \sp(2.0)
p_{\phi} = \sqrt{\gamma} n^a \partial_a \phi + \sqrt{\gamma} K \phi,
\cr
\sp(3.0)} \eqno(11)$$
$$ \eqalign{ \sp(2.0)
p_{\gamma} = {1 \over 4 \sqrt \gamma} n^a \partial_a (\phi^2),
\cr
\sp(3.0)} \eqno(12)$$
$$ \eqalign{ \sp(2.0)
p_A = - \phi^2 E.
\cr
\sp(3.0)} \eqno(13)$$
The Hamiltonian $H$ is expressed by a linear combination of constraints
as expected: 
$$ \eqalign{ \sp(2.0)
H = \int dx^0 ( \alpha H_0 + \beta H_1 + A_1 H_2 ), 
\cr
\sp(3.0)} \eqno(14)$$
where
$$ \eqalign{ \sp(2.0)
H_0 = {1 \over \gamma} [p_\Phi D_0 \Phi + {p_{\Phi^{\dag}}} (D_0 \Phi)^{\dag}]
 + {1 \over \gamma} p_\phi \dot \phi - 2 \dot p_\gamma - {1 \over 
\gamma} p_\gamma \dot \gamma, 
\cr
\sp(3.0)} \eqno(15)$$
$$ \eqalign{ \sp(2.0)
H_1 &= {1 \over {\sqrt{\gamma} \phi^2}} p_{\Phi} {p_{\Phi^{\dag}}} - 
{\sqrt {\gamma} \over 2} - {\dot \phi^2 \over {2 \sqrt{\gamma}}} 
+ \partial_0 ( {\partial_0 (\phi^2) \over {2 \sqrt{\gamma}}}) 
\cr
&\qquad+ {\phi^2 \over \sqrt{\gamma}} | D_0 \Phi |^2 - {2 \sqrt{\gamma} 
\over \phi} p_\phi p_\gamma + {2 \gamma \sqrt{\gamma} \over \phi^2} 
p_\gamma ^2  + {\sqrt{\gamma} \over {2 \phi^2}} p_A ^2 , 
\cr
\sp(3.0)} \eqno(16)$$
$$ \eqalign{ \sp(2.0)
H_2 = - ie (p_{\Phi} \Phi - p_{\Phi^{\dag}} \Phi^{\dag}) - \dot p_A.
\cr
\sp(3.0)} \eqno(17)$$
Note that $\alpha$, $\beta$ and $A_1$ are non-dynamical Lagrange 
multiplier fields.

The action can be cast into the first-order ADM canonical form by the 
dual Legendre transformation 
$$ \eqalign{ \sp(2.0)
S = \int d x^1 \bigl[ \int d x^0 ( p_{\Phi} \Phi^{\prime} + 
{p_{\Phi^{\dag}}} \Phi^{\prime \dag} 
+ p_{\phi} \phi^{\prime} + p_{\gamma} {\gamma}^{\prime} + p_A 
A_0^{\prime} ) - H \bigr]. 
\cr
\sp(3.0)} \eqno(18)$$
As Regge and Teitelboim pointed out [14], in order to have the correct 
Hamiltonian which produces the Einstein equations through the Hamilton 
equations, one has to supplement the surface terms to the Hamiltonian 
(14). In the present formalism, since we take the variation of all the
fields to be zero at both the singularity and the apparent horizon we 
do not have to add any surface terms to the Hamiltonian. This is a big 
difference between the present formalism considering the region 
inside the horizon and the 
previous formalism [11, 12] outside the horizon where a surface term at 
the spatial infinity needs to be added.

\vskip 1cm
\leftline{\bf 3. Quantum gravity between outer and inner apparent horizons }	
\par
In this section, for the purpose of applying the canonical formalism 
constructed in the previous section for understanding the mass inflation 
[4] in the Reissner-Nordstr$\ddot o$m black hole in quantum gravity, we 
will make a quantum gravity holding in the vicinity of apparent horizons 
where the complicated constraints have a tractable form while retaining 
the important features of the black hole physics. 
For simplicity, let us restrict ourselves to the neutral scalar field as 
the matter field, for which the $U(1)$ charge $Q$ has a fixed constant value 
and $p_A$ in Eq.(13) becomes $Q^2$. Thus under this situation the 
constraint $H_2$ generating the $U(1)$ gauge transformation identically 
vanishes. Moreover, we shall consider the charged Vaidya metric since the 
Reissner-Nordstr$\ddot o$m geometry is converted into the charged 
Vaidya form with a spherically symmetric stream of the massless matters. 

First of all, we shall consider the 
ingoing charged Vaidya metric.  In order to do so we will take the two 
dimensional coordinates $x^a$ by 
$$ \eqalign{ \sp(2.0)
x^a = (x^0, x^1) = (w - r, r),
\cr
\sp(3.0)} \eqno(19)$$
where $w$ is the standard external advanced time coordinate which is 
infinite both on the future null infinity and the Cauchy horizon [4, 7]. 
Note that we 
have chosen $x^1 = r$ as mentioned earlier. And we fix the 
gauge freedoms corresponding to the two dimensional 
reparametrization invariances by the gauge conditions
$$ \eqalign{ \sp(2.0)
g_{ab} &= \left(\matrix{ \gamma  & \alpha \cr
              \alpha & {\alpha^2 \over \gamma} - \beta^2 \cr} \right),
\cr
 &= \left(\matrix{ -(1 - {2M \over r} + {Q^2 \over {r^2}})  
            &  {2M \over r} - {Q^2 \over {r^2}} \cr
              {2M \over r} - {Q^2 \over {r^2}} 
            & 1 + {2M \over r} - {Q^2 \over {r^2}} \cr} \right),
\cr
\sp(3.0)} \eqno(20)$$
where $M$ is the mass function that depends only on $w$.
From these equations the two dimensional line element takes a form of the 
ingoing charged Vaidya metric
$$ \eqalign{ \sp(2.0)
ds^2 &= g_{ab} dx^a dx^b,
\cr
     &= -(1 - {2M \over r} + {Q^2 \over {r^2}}) dw^2 + 2 dw dr.
\cr
\sp(3.0)} \eqno(21)$$
Similarly, we can construct the outgoing charged Vaidya metric directly 
from 
the above ingoing one in the replacement of the advanced time coordinate 
by the retarded one. 

For later convenience let us introduce the two dimensional metric in 
radial double-null coordinates $(u, v)$, which has the form
$$ \eqalign{ \sp(2.0)
ds^2 = - 2 e^{- \lambda} du dv,
\cr
\sp(3.0)} \eqno(22)$$
where $\lambda = \lambda (u, v)$, and $u$ and $v$ are a retarded time and 
an advanced time, respectively.

For a dynamical black hole, it is sometimes useful to consider the local
definition 
of horizon, i.e., the apparent horizon, rather than the global one, the 
event horizon. In case of the Reissner-Nordstr$\ddot o$m black hole, 
the apparent horizon consists of two horizons. One is the outer apparent 
horizon given by $r_{+} =  M + \sqrt{M^2 - Q^2}$ and the other the 
inner apparent horizon $r_{-} =  M - \sqrt{M^2 - Q^2}$. In this article 
we confine ourselves to be $M > Q$.

In the literature [7], the classical analysis of the mass inflation 
has been achieved only on the two intersecting null surfaces $S^{+}$ and 
$S^{-}$ where the surfaces $S^{+}$ is radial rightgoing ($\nearrow$) null 
geodesics, and the surface $S^{-}$ radial leftgoing ($\nwarrow$) null 
one. We fix the 
system of coordinates such that $u =0$ is the event horizon and
$v =0$ the Cauchy horizon. Then we choose $S^{+}$ to be 
parallel to and near the event horizon while we 
take $S^{-}$ to coincide with the Cauchy horizon. 

Now we would like to consider the physics along $S^{+}$. According to 
a similar procedure to our previous work [9] which took advantage of 
the idea in the reference [15], let us attempt to solve the 
Hamiltonian and momentum constraints only in the vicinity of the 
outer apparent horizon. Here note that in the Poisson-Israel model [4], 
the outer apparent horizon approaches the event horizon asymptotically 
around the Cauchy horizon under a flow of infalling lightlike flux.
Thus it is reasonable to assume that $S^{+}$ is located in the vicinity 
of the outer apparent horizon since we are interested in the region $S = 
S^{+} \cap S^{-}$.
Near the outer apparent horizon, we make an 
approximation 
$$ \eqalign{ \sp(2.0)
\Phi \approx \Phi(w), M \approx M(w), \phi \approx r.
\cr
\sp(3.0)} \eqno(23)$$
From now on we shall use $\approx$ to indicate the equalities which hold 
approximately near the apparent horizons. Indeed one can prove the above 
assumptions (23) to be consistent with the field equations [9].

Near the outer apparent horizon $r_{+}$, Eq.(20) yields
$$ \eqalign{ \sp(2.0)
\alpha \approx +1 , \gamma = {1 \over \beta^2} \approx 0,
\cr
\sp(3.0)} \eqno(24)$$
and the canonical conjugate momenta (10)-(12) are given 
approximately as
$$ \eqalign{ \sp(2.0)
p_{\Phi} \approx - \phi^2 \partial_w \Phi,
\cr
p_{\phi} \approx - {1 \over \gamma} \partial_w M,
\cr
p_{\gamma} \approx -{1 \over 2} \phi .
\cr
\sp(3.0)} \eqno(25)$$
Then it is easy to check the fact that the two constraints are 
proportional to each other
$$ \eqalign{ \sp(2.0)
- \gamma H_0 &\approx \sqrt{\gamma} H_1,
\cr
             &\approx {1 \over \phi^2} p_{\Phi}^2 + \gamma p_\phi. 
\cr
\sp(3.0)} \eqno(26)$$
As in the previous work [9], just at the apparent horizon $\gamma$ 
becomes zero so that the various equalities approximately hold when
we restrict our attention to the interior region 
near but not at the apparent horizon. Thus we should assume that 
$\gamma$ takes a small but finite value within the present approximation 
level. 

An imposition of the constraint (26) as an operator equation on the state 
produces the Wheeler-DeWitt equation
$$ \eqalign{ \sp(2.0)
i {\partial \Psi \over {\partial \phi}} = - {1 \over {\gamma 
\phi^2}} {\partial^2 \over {\partial \Phi^2}} \Psi.
\cr
\sp(3.0)} \eqno(27)$$
This Wheeler-DeWitt equation is regarded as the Schr$\ddot o$dinger 
equation in the superspace with the Hamiltonian $H_S = p_{\Phi}^2$ 
and the time $T_S = {1 \over {\gamma \phi}}$. Note that this 
superspace time $T_S$ stops on the horizon owing to the effect of an 
infinite gravitational time dilation.

Now it is easy to find a special solution of this Wheeler-DeWitt 
equation by the method of separation of variables. The result is 
$$ \eqalign{ \sp(2.0)
\Psi = (B_w e^{\sqrt A_w \Phi(w)} + C_w e^{-\sqrt A_w \Phi(w)} ) 
        \ e^{- i{A_w \over {\gamma \phi}} },
\cr
\sp(3.0)} \eqno(28)$$
where $A_w$, $B_w$, and $C_w$ are integration constants. 

The formalism explained so far is easily extendible to the quantum 
gravity near the inner apparent horizon. The result is nothing but the 
replacement of the advanced time variable $w$ by the retarded one $z$. 
For later convenience, we will write down the conjugate momenta and 
the physical state in the below:
$$ \eqalign{ \sp(2.0)
p_{\Phi} &\approx - \phi^2 \partial_z \Phi,
\cr
p_{\phi} &\approx - {1 \over \gamma} \partial_z M,
\cr
p_{\gamma} &\approx -{1 \over 2} \phi,
\cr
\Psi &= (B_z e^{\sqrt A_z \Phi(z)} + C_z e^{-\sqrt A_z \Phi(z)} ) 
        \ e^{- i{A_z \over {\gamma \phi}} },
\cr
\sp(3.0)} \eqno(29)$$
where $A_z$, $B_z$, and $C_z$ are integration constants, and the matter 
field $\Phi$ and the mass function $M$ were assumed to be the 
function of only the standard external retarded time coordinate 
$z$ near the inner apparent 
horizon.

At this stage, if one defines an expectation value $< \cal O >$ 
of an operator $\cal O$ rather naively as
$$ \eqalign{ \sp(2.0)
< {\cal O} > = {1 \over {\int d\Phi |\Psi|^2 }} \int d\Phi \Psi^* {\cal O} 
\Psi,
\cr
\sp(3.0)} \eqno(30)$$
one can calculate $< \partial_w M >$ by using either (25) or (26)
$$ \eqalign{ \sp(2.0)
< \partial_w M > = - {A_w \over  {< \phi^2 >} }.
\cr
\sp(3.0)} \eqno(31)$$
This equation shows the black hole radiation when one chooses the 
constant $A_w$ to be a positive constant $k_{w}^2$. Then the mass-loss rate 
becomes [9]
$$ \eqalign{ \sp(2.0)
< \partial_w M > = - {k^2 \over {< r_{+}^2 >}}.
\cr
\sp(3.0)} \eqno(32)$$
If we consider the Schwarzschild black hole, this result exactly 
corresponds to that calculated by Hawking in the 
semiclassical approach [10] and by Tomimatsu in the exterior region of 
the apparent horizon in quantum gravity [15]. Thus a black hole 
completely evaporates within a finite time. However, we are now thinking 
of the outer apparent horizon of the Reissner-Nordstr$\ddot o$m 
black hole, for which if we apply the Stefan-Boltzmann law of the 
blackbody radiation by assuming the radiation area to be that of the 
outer horizon and the temperature to 
be the Hawking temperature [16] 
in the Reissner-Nordstr$\ddot o$m black hole,  
we meet an inconformity, for which we will not pay attention in this article.

\vskip 1cm
\leftline{\bf 4. Mass inflation in quantum gravity }	
\par
The classical physics behind the mass inflation is not so difficult to 
understand and arises from two key observations. One is that the influx 
is tremendously blueshifted along the Cauchy horizon so that the energy 
density of the influx inflates exponentially. The other is a 
separation between the Cauchy horizon, a surface of infinite blueshift, 
and inner apparent horizon, a surface of infinite redshift, owing to 
some transverse outflux. As a consequence, the mass function near the 
Cauchy horizon diverges 
exponentially with respect to an advanced time.

Now let us proceed with the study of the mass inflation in quantum 
gravity. According to the literature [7], let us analyse this phenomenon 
only on the two intersecting null surfaces $S^{+}$ and $S^{-}$. As mentioned 
before, we will choose $S^{-}$ to coincide with the Cauchy horizon, and  
$S^{+}$ to be parallel to as well as near the 
event horizon. 

As a preparation, let us introduce the extrinsic fields [7]
$$ \eqalign{ \sp(2.0)
\theta := \phi^{-2} \partial_v (\phi^2),
\cr
\tilde \theta := \phi^{-2} \partial_u (\phi^2),
\cr
\sp(3.0)} \eqno(33)$$
where $\theta$ and $\tilde \theta$ denote the expansion rate of the 
ingoing and the outgoing fluxes, respectively. Along $S^{+}$ the radius 
satisfies the equation 
$$ \eqalign{ \sp(2.0)
{dr \over {dw}}= {1 \over 2} \bigl[ 1 - {2M(w) \over r} + {Q^2 \over {r^2}} 
\bigr].
\cr
\sp(3.0)} \eqno(34)$$
Then we can make 
a choice of $\lambda = \log \phi$ along $S^{+}$ by means of the freedom 
of the rescaling of the null coordinates. By solving a focusing equation [7] 
and Eq.(34), it is relatively straightforward to find an explicit
relation between $v$ and $w$ near the Cauchy horizon:
$$ \eqalign{ \sp(2.0)
v \approx - e^{- \kappa_0 w},
\cr
\sp(3.0)} \eqno(35)$$
where $\kappa_0$ is the surface gravity of the inner horizon. Then since 
the expansion rate of the 
ingoing flux in Eq.(33) along $S^{+}$ near the Cauchy horizon $v 
\rightarrow 0$ ($w \rightarrow \infty$) can be expressed as 
$$ \eqalign{ \sp(2.0)
\theta :&= \phi^{-2} \partial_v (\phi^2) \approx r^{-2} \partial_v (r^2)
\cr
        &\approx - {2 \over {\kappa_0 v \sqrt{M^2 - Q^2}}} \partial_w M
        \approx {2 \gamma \over {\kappa_0 v \sqrt{M^2 - Q^2}}} p_{\phi},
\cr
\sp(3.0)} \eqno(36)$$
we can easily evaluate the expectation value of which  
$$ \eqalign{ \sp(2.0)
< \theta >_{S^{+}} \approx  { 2 A_w \over {\kappa_0 v \sqrt{M^2 - Q^2} < 
r_{+}^2 > } }.
\cr
\sp(3.0)} \eqno(37)$$
In the present physical setting, the influx of the massless neutral 
matter $\Phi(w)$ flows into the black hole through $r = r_{+}$ so that $< 
\partial_w M >$ must be positive, which requires the constant $A_w$ in 
Eq.(31) to be a certain negative constant, e.g., $- k_{w}^2$, which 
should be contrasted with the case of the Hawking radiation considered in 
the previous section, where the constant $A_w$ was selected to be a 
positive constant. 
Consequently, Eq.(37) becomes 
$$ \eqalign{ \sp(2.0)
< \theta >_{S^{+}} \approx  - { 2 k_{w}^2 \over {\kappa_0 v 
\sqrt{M^2 - Q^2} < r_{+}^2 > } },
\cr
\sp(3.0)} \eqno(38)$$
as $v \rightarrow 0$. This result indicates that, while the radius of the 
two-sphere $S = S^{+} \cap S^{-}$ is finite, $< \theta >_{S^{+}} 
\rightarrow -\infty$ at the Cauchy horizon, which is a signal of the mass 
inflation. It is worth remarking here 
that the divergent behavior of the expansion rate comes from a huge 
blueshift factor $1 \over v$. This situation is completely 
analogous to the classical result [7] although the way of deriving those 
results are very much different in both methods. 

To show the mass inflation more clearly, let us consider the invariant 
definition of the black hole mass in a spherically symmetric geometry 
$$ \eqalign{ \sp(2.0)
1 - { 2 M(x^a) \over r} + {Q^2 \over r^2} := g^{ab} \partial_a r 
\partial_b r  \approx  - { r^2 \over 2} e^{\lambda} \theta \tilde \theta,
\cr
\sp(3.0)} \eqno(39)$$
where 
$$ \eqalign{ \sp(2.0)
\theta := \phi^{-2} \partial_v (\phi^2) \approx r^{-2} \partial_v (r^2),
\cr
\tilde \theta := \phi^{-2} \partial_u (\phi^2) \approx r^{-2} \partial_u
 (r^2).
\cr
\sp(3.0)} \eqno(40)$$
Now on the two-sphere $S = S^{+} \cap S^{-}$ we are interested in 
calculating an expectation value of Eq.(39). In order to do so, one 
needs to calculate $<\tilde \theta>$ and $<\partial_u M>$ on $S^{-}$. 
In the model considered at present, the outflux is switched on at some 
retarded time $u = u_0 \approx 0$. Since near $u = u_0$ the inner 
apparent horizon almost coincides with the Cauchy horizon, we can make 
use of the formalism holding the inner apparent horizon (29). Thus, 
similar calculations 
to those of $<\theta>_{S^{+}}$ and $<\partial_v M>_{S^{+}}$ yield
$$ \eqalign{ \sp(2.0)
<\tilde \theta>_{S^{-}}  \approx  { 2 A_z \over {\kappa_0 u \sqrt{M^2 - Q^2} < 
r_{-}^2 > } },
\cr
\sp(3.0)} \eqno(41)$$
$$ \eqalign{ \sp(2.0)
< \partial_u M >_{S^{-}} \approx - {A_z \over {< r_{-}^2 >}}.
\cr
\sp(3.0)} \eqno(42)$$
We can interpret that the outflux of lightlike radiation through $r = 
r_{-}$ from the $r_{-} < r < r_{+}$ region to $0 < r < r_{-}$ region 
comes from the surface of the collapsing star [4], for which we must 
take the constant $A_z$ to be a negative constant, e.g., $- k_z^2$. As a 
result, after some manipulation, we obtain 
$$ \eqalign{ \sp(2.0)
<1 - { {2 M(x^a)} \over r} + {Q^2 \over r^2}>_{S = S^{+} \cap S^{-}} 
\approx   - {  2 k_w^2 k_z^2 \over {\kappa_0^2 v u (M^2 - Q^2) 
 \sqrt {< r_{+} r_{-} >} } }. 
\cr
\sp(3.0)} \eqno(43)$$
Note that the right-hand side of this equation is actually a negative 
infinity near the Cauchy horizon, which implies that $< M > \rightarrow 
\infty$, that is, $\it mass \ inflation$. The 
important point here is that the mechanism responsible for the mass 
inflation needs the combination of the effects of both the outflux and 
the influx. Incidentally we need not fear that the right-hand side of 
Eq.(43) also becomes divergent at the event horizon $u=0$ since there is 
no outflux before $u = u_0 > 0$.
This situation is also perfectly analogous to the classical 
result [4].

Finally, the square of the Weyl tensor on the  $S = S^{+} \cap S^{-}$ can 
be calculated to be
$$ \eqalign{ \sp(2.0)
<C_{\alpha \beta \gamma \delta} C^{\alpha \beta \gamma \delta}>_S 
&= 12 \ \bigl( {1 \over {r^2}} - {Q^2 \over {r^4}} + {r \over 2} 
\theta \tilde 
\theta \bigr) ^2  
\cr
&\propto <\theta^2>_{S^{+}}  <\tilde \theta^2>_{S^{-}}
\cr
&\propto  + {1 \over v^2}, 
\cr
\sp(3.0)} \eqno(44)$$
from which we can imagine that the square of the Weyl tensor on the 
Cauchy horizon diverges rather strongly, thus a quantum continuation 
beyond the Cauchy horizon seems to be unlikely although we need a more
detailed analysis to confirm this statement.

\vskip 1cm
\leftline{\bf 5. Conclusion }	
\par
In this article, we have explored the mass inflation which is the 
phenomenon that the local mass function reaches an incredibly large value 
in the interior of the charged black hole, by using the canonical 
quantization formalisms of gravity developed in our previous work [9]. As 
pointed out in the pioneering work by Poisson and Israel [4], the mass 
inflation depends on the general feature, that is, the presence of 
a highly blueshifted influx and a small outflux. It is remarkable to 
notice that this feature in the classical theory of general relativity 
is inherited to the case of quantum gravity. 

In the sec.$V$ in the reference [4], Poisson and Israel have speculated 
that the effects of quantum gravity might not alter the classical picture 
of the mass inflation. At least our present results seem to support their 
speculations. It would be very interesting to show in more general 
context that the quantum solutions should be still singular if the 
solutions to the classical equation all exhibit a singular behavior [4]. 
(However, see the recent work which seems to be against this conjecture 
[17].)  It might be not quantum gravity but the combined effects of 
quantum gravity and supersymmetry that prevent an appearance of the 
classical singular solutions, which was shown at least in the context 
of the two dimensional dilaton gravity [18]. 

Kuchar has recently constructed an interesting canonical quantization 
formalism based on the Kruskal extention of the Schwarzschild black hole 
[19]. 
In his work, the hypersurfaces which foliate the spacetime extend from 
the left spatial infinity to the right one in the Kruskal diagram, thus 
it would be very interesting to 
investigate the mass inflation by his formalism.

In future works, we would like to investigate further other recently
developing problems such as the smearing of the black 
hole singularities [20, 21] and the black hole thermodynamics [22, 23] by 
using the canonical formalism holding the interior region of black hole.

\vskip 1cm

\noindent

\leftline{\bf Acknowledgments}

The author is indebted to Professor A.Hosoya for stimulating discussions 
and collaboration at early stage of this work. He would also like to 
thank Professor E.Elizalde for a kind hospitality at Barcelona University
in Spain where a part of the present work has been completed.

\vskip 1cm

\noindent

\leftline{\bf References}
\par
\item{[1]} R.M.Wald, General Relativity (The University of Chicago Press, 
1984). 

\item{[2]} R.Penrose, Riv. del Nuovo Cim.(numero speziale) {\bf 1} (1969) 
252.

\item{[3]} S.Chandrasekhar, The Mathematical Theory of Black Holes ( 
Oxford University Press, 1983) and references therein.  

\item{[4]} E.Poisson and W.Israel, Phys. Rev. {\bf D41} (1990) 1796.

\item{[5]} A.Ori, Phys. Rev. Lett. {\bf 67} (1991) 789.

\item{[6]} A.Ori, Phys. Rev. Lett. {\bf 68} (1992) 2117.

\item{[7]} P.R.Brady and C.M.Chambers, Phys. Rev. {\bf D51} (1995) 4177.

\item{[8]} P.R.Brady and J.D.Smith, Phys. Rev. Lett. {\bf 75} (1995) 1256.

\item{[9]} A.Hosoya and I.Oda, ``Black Hole Radiation inside Apparent 
Horizon in Quantum Gravity'', EDO-EP-7, gr-qc/9610055.

\item{[10]} S.W.Hawking, Comm. Math. Phys. {\bf 43} (1975) 199.

\item{[11]} P.Thomi, B.Isaak, and P.Hajicek, Phys. Rev. {\bf D30} (1984) 
1168.

\item{[12]} P.Hajicek, Phys. Rev. {\bf D30} (1984) 1178.

\item{[13]} C.W.Misner, K.S.Thorne, and J.A.Wheeler, Gravitation (Freeman, 
1973).

\item{[14]} T.Regge and C.Teitelboim, Ann. Phys. (N.Y.) {\bf 88} (1974) 286.

\item{[15]} A.Tomimatsu, Phys. Lett. {\bf B289} (1992) 283.

\item{[16]} G.W.Gibbons and S.W.Hawking, Phys. Rev. {\bf D15} (1977) 2738.

\item{[17]} A.Ashtekar, ``Large Quantum Gravity Effects: Unexpected 
Limitations of the Classical Theory'', gr-qc/9610008 .

\item{[18]} S.Nojiri and I.Oda, Mod. Phys. Lett. {\bf A8} (1993) 53.

\item{[19]} K.V.Kuchar, Phys. Rev. {\bf D50} (1994) 3961.

\item{[20]} A.Hosoya, Class. Quant. Grav. {\bf 12} (1995) 2967.

\item{[21]} A.Hosoya and I.Oda, ``Black Hole Singularity and Generalized 
Quantum Affine Parameter'', TIT/HEP-334/COSMO-73, EDO-EP-5, gr-qc/9605069.

\item{[22]} S.Carlip, ``The Statistical Mechanics of Horizons and Black 
Hole Thermodynamics'', UCD-96-10, gr-qc/9603049; ``The Statistical 
Mechanics of the Three-Dimensional Euclidean Black Hole'', UCD-96-13, 
gr-qc/9606043. 

\item{[23]} C.Teitelboim, Phys. Rev. {\bf D53} (1996) 2870.

\endpage
%

%
\bye